\documentstyle[12pt,aasms4,psfig]{article}
\evensidemargin=\oddsidemargin

\def\degr{\hbox{$^\circ$}}

\def\farcm{\hbox{$.\mkern-4mu^\prime$}}
\def\farcs{\hbox{$.\!\!^{\prime\prime}$}}         
\def\gsim{\mathrel{\hbox{\rlap{\lower.55ex \hbox {$\sim$}}
                   \kern-.3em \raise.4ex \hbox{$>$}}}}

\begin{document}
\begin{center}

{\bf UNDER EMBARGO AT NATURE (submitted June 10, 1998)}

\vspace{1cm}

{\bf Discovery of the peculiar supernova 1998bw in the 
error box of GRB980425}

{T.J. Galama$^1$, P.M. Vreeswijk$^1$, J. van Paradijs$^{1,2}$, 
C. Kouveliotou$^{3,4}$,
T. Augusteijn$^5$, O.R. Hainaut$^5$, F. Patat$^5$, 
H. B\"ohnhardt$^5$, J. Brewer$^5$, V. Doublier$^5$,
J.-F. Gonzalez$^5$, C. Lidman$^5$, B. Leibundgut$^{5}$, 
J. Heise$^6$, J. in 't Zand$^6$, P.J. Groot$^1$, R.G. Strom$^{1,7}$, 
P. Mazzali $^{8}$, K. Iwamoto$^{9}$, K. Nomoto$^{9,10}$, H. Umeda$^{9,10}$,
T. Nakamura$^{9}$, T. Koshut$^{3,4}$, M. Kippen$^{3,4}$, 
C. Robinson$^{3,4}$, P. de Wildt$^{1}$, R.A.M.J. 
Wijers$^{11}$, N. Tanvir$^{11}$, J. Greiner$^{12}$, 
E. Pian$^{13}$, E. Palazzi$^{13}$, F. Frontera$^{13}$, 
N. Masetti$^{13}$, L. Nicastro$^{13}$, E. Malozzi$^{13}$, 
M. Feroci$^{14}$, E. Costa$^{14}$, L. Piro$^{14}$, B.A. Peterson$^{15}$,
C. Tinney$^{16}$, B. Boyle$^{16}$, R. Cannon$^{16}$, R. Stathakis$^{16}$, 
M.C. Begam$^{17}$, P. Ianna$^{17}$}

\end{center}

\noindent $^1$ Astronomical Institute ``Anton Pannekoek'', University
of Amsterdam, \& Center for High Energy Astrophysics, Kruislaan 403,
1098 SJ Amsterdam, The Netherlands 
\newline $^2$ Physics Department,
University of Alabama in Huntsville, Huntsville, AL 35899, USA
\newline $^3$ Universities Space Research Association 
\newline $^4$
NASA Marshall Space Flight Center, ES-84, Huntsville, AL 35812, USA
\newline $^5$ ESO, Casilla 19001, Santiago 19, Chile
\newline $^6$ SRON Laboratory for
Space Research, Sorbonnelaan 2, 3584 CA Utrecht, The Netherlands
\newline $^{7}$ Netherlands Foundation for Research in Astronomy,
Postbus 2, 7990 AA Dwingeloo, The Netherlands 
\newline $^{8}$ Osservatorio Astronomico di Trieste, Via G.B. Tiepolo 11, 
I-34131 Trieste, Italy
\newline $^{9}$ Department of Astronomy, School of Science, University of 
Tokyo, Tokyo 111, Japan
\newline $^{10}$ Research center for the early Universe, School of Science, 
University of Tokyo, Tokyo 111, Japan
\newline $^{11}$ Institute of Astronomy, Madingley Road, Cambridge CB3 0HA, UK 
\newline $^{12}$ Astrophysikalisches Institut, Potsdam, Germany 
\newline $^{13}$ Instituto Tecnologie e Studio Radiazioni Extraterrestri, 
CNR, Bologna, Italy
\newline $^{14}$ Instituto di Astrofisica Spaziale, CNR, Roma, Italy 
\newline $^{15}$ Mt. Stromlo and Siding Spring Observatories, The Australian
National University, Weston Creek A. C. T. 2611, Australia
\newline $^{16}$ Anglo-Australian Observatory, PO Box 296, Epping, NSW 2121, Australia
\newline $^{17}$ Dept. of Astronomy, University of Virginia,
Charlottesville, VA 22903, PO Box 3818, USA
\newpage

{\bf The discovery of X-ray$^{1}$, optical$^{2}$ and radio$^3$
afterglows of $\gamma$-ray bursts (GRBs) and the measurements of the
distances to some of them$^{4,5}$ have established that these events
come from Gpc distances and are the most powerful photon emitters
known in the Universe, with peak luminosities up to $10^{52}$
erg\,s$^{-1}$. We here report the discovery of an optical transient,
in the BeppoSAX Wide Field Camera error box of GRB\, 980425, which
occurred within about a day of the $\gamma$-ray burst. Its optical
light curve, spectrum and location in a spiral arm of the galaxy
ESO\,184-G82, at a redshift $z = 0.0085^{6}$, show that the transient
is a very luminous type Ic supernova, SN\,1998bw. The peculiar nature
of SN\,1998bw is emphasized by its extraordinary radio properties
which require that the radio emitter expand at relativistical
speed$^{7,8}$.  Since SN\,1998bw is very different from all previously
observed afterglows of GRBs, our discovery raises the possibility that
very different mechanisms may give rise to GRBs, which differ little
in their $\gamma$-ray properties.  }

GRB\,980425 was detected$^{9,10}$ on April 25.90915 UT with one of the
Wide Field Cameras (WFCs) and the Gamma Ray Burst Monitor (GRBM) on
board BeppoSAX, and with the Burst and Transient Source Experiment
(BATSE) on board the Compton Gamma Ray Observatory (CGRO).

The BATSE burst profile consists of a single wide peak. 
The burst flux rose in $\sim 5$ s to a maximum flux of $(3.0\pm 
0.3) \times 10^{-7}$ erg \,cm$^{-2}$\,s$^{-1}$ (24 --1820 keV), at which it 
remained for $\sim 5$ s; it decayed steadily to the background 
in $\sim 25$ s. The burst fluence $E_{\rm b} = (4.4 \pm 0.4) \times 10^{-
6}$ erg\,cm$^{-2}$; its duration$^{11}$ $T_{90} = 23.3 \pm 1.4$ s. No
burst emission is detected above 300 keV; GRB\,980425 is therefore a
so-called NHE burst$^{12}$, a group comprising about a quarter 
of the GRBs detected with BATSE. The 
burst spectrum is well described by a smoothly broken power law, 
with a constant break energy ($148 \pm 33$ keV) and high-energy power law 
photon index ($-3.8 \pm 0.7$); the low-energy power law photon index
varied from  
$-1.0 \pm 0.15$ during the rise, to $-2.6 \pm 0.2$ during the decay of the 
burst (i.e., the burst spectrum became softer). These results show that 
with respect to duration, $\gamma$-ray 
spectrum, peak flux and fluence, GRB\,980425 was not a remarkable event.

In the BeppoSAX WFC no. 2 the burst lasted $\sim 30$ a, 
and reached a peak intensity of about 3 Crab (2-28
keV)$^9$. The position derived from the WFC image is RA= 19$^{\rm
h}34^{\rm m}54^{\rm s}$, Decl = --52\degr49\farcm9 (equinox 2000.0),
with an error radius of 8$^\prime$ which comprises a 3$^\prime$ 
statistical error (99\% confidence level) and a 5$^\prime$ 
systematic uncertainty due to incomplete
satellite attitude information$^{13,14}$. 

We observed the error box of GRB\,980425 in$^{15}$ R$_{\rm Macho}$ and
B$_{\rm Macho}$ with the 30 and 50 inch telecopes at the Australian
National University's (ANU) Mt. Stromlo Observatory (MSO) starting
April 26.60 UT, and in standard (U, B, V, R and I) with the 40 inch
telescope at the ANU Siding Spring Observatory (SSO), the 3.5m New
Technology Telescope (NTT), and the 1.5m Danish and the 0.9m Dutch
telescopes at the European Southern Observatory (ESO).

Inspection of NTT images obtained on Apr. 28.4 and May 1.3 UT revealed
a point source in the WFC error box, which was not visible in the
Digitized Sky Survey$^{16}$ (see Fig. 1).  Using the US Naval
Observatory catalog and the NTT May 1.3 UT R band image we determine
its position at R.A. = $19^{\rm h}35^{\rm m}03.14^{\rm s} \pm
0.12^{\rm s}$, Decl. = $-52^{\circ}50^{\prime}45.3^{\prime\prime} \pm
1.0^{\prime \prime}$ (equinox 2000.0), 1\farcm6 away from the center
of the WFC error box. The source coincides with the transient radio
source in the WFC error box$^7$ to within 1\farcs8 (i.e.,
$1.2\sigma$), and it is located in a spiral arm of the face-on barred
spiral galaxy ESO 184-G82 at a redshift of $z = 0.0085^{6}$, in the DN
1931--529 group of galaxies$^{17}$.

The UBVRI light curves of SN\,1998bw are shown in Fig. 2. The R band
light curve shows an initial `plateau', then it rises at a rate of
$\sim 0.25$ mag day$^{-1}$ to maximum light, after which it declines
by $\sim 0.05$ mag day$^{-1}$.  The plateau may be the late signature
of a sharp initial peak in the light curve, signalling the shock break
out at the surface of the progenitor star$^{18}$.  Lack of early data
prevents us from establishing its existence in the U, B, V and I light
curves.

Times of maximum, and peak magnitudes in the five bands are listed in
Table 1, together with the peak absolute magnitudes (we used H$_0$ =
65 km s$^{-1}$ Mpc$^{-1}$, a redshift $z = 0.0085$, and corrected for
galactic foreground extinction, $A_V$ = 0.20, as inferred from a
combination of COBE/DIRBE and IRAS/ISSA maps$^{19}$).  The longer the
wavelength is, the later maximum light occurs.

H and He lines are absent in the early spectra of SN\,1998bw, which
exclude SN types II and Ib, respectively. Type Ia supernovae are
normally identified by a Si II line, which is not detected in the
spectrum of SN\,1998bw. This has led to the classification of this
event as a type Ic supernova$^{16,20}$.  The light curve of SN\,1998bw
is inconsistent with a type Ia, although the luminosity of SN\,1998bw
rivals that of typical type Ia supernovae. To achieve such
luminosities substantial amounts of $^{56}$Ni ($\sim 0.7$\,
M$_{\odot}$) have to be synthesized in the explosion$^{21}$, which
would be unprecedented for a core-collapse supernova.  (Typical values
for the amount of $^{56}$Ni are $< 0.1$\,M$_{\odot}$; the most
luminous SN\,II event so far has been SN\,1992am with a $^{56}$Ni mass
of 0.3\,M$_{\odot}$ $^{22}$.

Modelling$^{21}$ of the optical light curve of SN\,1998bw shows that
it can be reproduced with the core collapse of a massive CO progenitor
star; the time of collapse coincides with that of the GRB to within
about a day. In the case of the CO star core collapse the kinetic
energy was $\sim 10^{52.5}$ ergs; the total energy (including
neutrinos) likely exceeded $10^{54}$ ergs.

Any estimate of the probability that the supernova and the GRB
coincided by chance (both with respect to time and direction) suffers
from the problem of {\it a posteriori} statistics, i.e., that the
parameters of the problem tend to be set by the observed phenomenon
itself. In this case the parameters are the size of the error box, the
peak magnitude of the supernova, and the time window within which the
events can be considered as possibly related. In our estimate we have
made generous estimates of these parameters.

The WFC error boxes have 99\% confidence level radii varying between
3$^\prime$ and 8$^\prime$ (Ref. 13). We conservatively estimate the
angular distance beyond which a connection can be rejected, at
10$^\prime$. We included all supernovae with peak magnitudes $m_B
<16$, i.e., $\sim 2$ mag below that of SN1998bw.  The time of
occurrence of the GRB and the onset of the supernova coincide to
within about a day (see above). Since a GRB which would have occurred
a few days earlier or later would have been considered at least
remarkable, we have taken a time window of 10 days.

With peak absolute magnitudes $M_B = -18.28, -16.68,$ and $-15.69$ for
supernovae of types Ia, Ib/c and II, respectively$^{23}$, for $m_B
<16$ they are detectable out to redshifts of 7180, 3440 and 2180 km/s,
respectively (note that the limiting values of $z$ are independent of
the assumed value of the Hubble constant).  The Shapley-Ames
`fiducial' sample of 342 galaxies within the Virgo circle$^{23}$ has a
mean B-band luminosity of $6.7 \times 10^{9}$ L$_{\odot}(B)$, and a
supernova rate of $3.09\, [100\, {\rm yr}\, 10^{10} {\rm
L}_{\odot}(B)]^{-1}$. Using galaxy numbers and heliocentric radial
velocities from Ref. 24, assuming a mean luminosity, galaxy
composition, and SN rate as in the `fiducial' sample, and taking
relative supernova rates$^{25}$ for types II: Ib/c : Ia = 4: 2: 1, we
find a total rate of supernovae (with $m_B < 16$ at the peak) of 120
per year. This value includes a correction for absorption within the
host galaxy disk. This number should perhaps be increased by a modest
factor to account for incompleteness of the radial velocity
distribution$^{24}$; we have adopted a final SN rate ($m_B <16$) of
150 per year.

With the above parameters we estimate the probability of catching a SN
in one of the 13 WFC GRB error boxes to be $1.1 \times 10^{-4}$. In
our probability estimate we have included all supernovae with peak
magnitudes two magnitudes below that of SN\,1998bw, and we have
ignored the fact that SN\,1998bw is of a rare type. We therefore
believe our estimate is conservative.  As a result, the notion that
GRB\,980425 and SN\,1998bw are physically related becomes difficult to
reject purely on the basis of the fact that afterglows observed so far
from GRB are very different from supernovae.

Follow-up X-ray observations with the BeppoSAX narrow-field
instruments, showed that the WFC error box contains two X-ray
sources$^{26,27}$, neither of which coincides with SN\,1998bw.  One,
1SAX\,J1935.0--5248 has a constant (2--10 keV) flux of $\sim 2 \times
10^{-13}$ erg \,cm$^{-2}$\,s$^{-1}$. The other, 1SAX\,J1935.3--5252,
was detected at $(1.6 \pm 0.3) \times 10^{-13}$ erg
\,cm$^{-2}$\,s$^{-1}$ about 1 day after the burst, and decayed to
$<1.2 \times 10^{-13}$ erg \,cm$^{-2}$\,s$^{-1}$ ($3\sigma$) in 22
hours; it was not detected 6 days after the burst ($< 1.0 \times
10^{-13}$ erg \,cm$^{-2}$\,s$^{-1}$).  This variability is similar to
that of previously observed X-ray afterglows of GRBs, and this object
might be a possible counterpart for GRB\,980425. Comparison of the
Mt. Stromlo April 26.63 UT and Apr 28.68 UT R$_{\rm Macho}$ band
images at the locations of the two X-ray sources shows no variation
$>$ 0.2 mag down to 21 mag$^{28,29}$. However, in several cases
optical afterglows were not detected, most notably GRB\,970111$^{30}$
and GRB\,970828$^{31}$.

The (2-10) keV detection limit ($3\sigma$) for the GRB 980425 NFI
observations was $1.2 \times$ 10$^{-13}$ erg s$^{-1}$cm$^{-2}$. Using
the ASCA (2-10 keV) source count distributions$^{33}$ one expects to
find an average of 0.6 X-ray sources above this limit in the WFC error
box; the probability of finding two or more sources there by chance
coincidence is 12\%. The case for a relation between this X-ray source
and GRB\,980425 must therefore be considered tentative at best, in
particular since among weak ROSAT sources variability is not
rare$^{32}$.

If one accepts the possibility that GRB 980425 and SN\,1998bw are
associated, one must conclude that GRB\,980425 is a rare type of GRB,
and SN\,1998bw is a rare type of supernova. The optical and radio
properties of SN\,1998bw show the rare nature of this event
independent of whether or not it is associated with GRB\,980425.

The consequence of an association is that the $\gamma$-ray peak
luminosity of GRB 980425 is L$_{\gamma}$ = 5.5 $\pm$ 0.7 $\times
10^{46}$ erg \,s$^{-1}$ (24 --1820 keV) and its total $\gamma$-ray
energy budget of 8.5 $\pm$ 1.0 $\times 10^{47}$ erg. These values are
much smaller than those of `normal' GRBs wich have peak luminosities
of up to $10^{52}$ erg \,s$^{-1}$ and a total energy budget of several
times $10^{53}$ erg$^{5}$.  This implies that very different
mechanisms can produce GRBs which cannot be distinguished on the basis
of their $\gamma$-ray properties.

The radio and optical properties of SN\,1998bw imply the simultaneous
presence of a photosphere moving out at several $10^4$ km/s, and a
relativistic outflow (this accounts for the super-Compton brightness
temperature of the radio source$^8$, and is required for the
$\gamma$-ray burst emission).  The relativistic flow could perhaps
develop in the medium around the supernova after shock
breakout. Alternatively, it may require anisotropy of the SN
explosion, possibly related to rapid rotation. The latter picture is
reminiscent of the `hypernova' model, i.e., a very massive core
collapse to a black hole, temporarily encircled by a disk, proposed by
Paczy\'nski$^{34}$ as a mechanism for GRBs.  If SN\,1998bw is caused
by such a 'hypernova', this model is unlikely to apply to the 'normal'
GRBs, which are $\sim 10^5$ times more energetic in $\gamma$ rays than
GRB\,980425.

This work is based partly upon images obtained by the MACHO Project
with the 50in telescope and the RAPT Group with the 30in telecope at
the Australian National University's Mt. Stromlo Observatory, and by
Dr. H. Jerjen with the 40in telescope at the Australian National
University's Siding Spring Observatory. RMK and KH acknowledge support
from NASA grant NAG5-6747.

\newpage

\references

1. Costa, E. {\it et al.} Discovery of an X-ray afterglow associated
with the gamma-ray burst of 28 February 1997. {\it Nature} {\bf 387},
783-785 (1997).\\
2. Van Paradijs, J. {\it et al.} Transient optical
emission from the error box of the $\gamma$-ray burst of 28 February
1997. {\it Nature} {\bf 368}, 686-688 (1997).\\ 
3. Frail, D.A., Kulkarni, S.R., Nicastro. L., Feroci, M., Taylor, G.B.
The radio afterglow from the $\gamma$-ray burst of 8 May 1997. {\it
Nature} {\bf 389}, 261-263 (1997) .\\
4. Metzger, M.R. {\it et al.} Spectral constraints on the redshift of
the optical counterpart to the gamma-ray burst of 8 May 1997. {\it
Nature} {\bf 387}, 878-880 (1997) .\\
5. Kulkarni, S.R. {\it et al.} Identification of a host galaxy at
redshift z = 3.42 for the $\gamma$-ray burst of December 1997. {\it
Nature} {\bf 393}, 35-39 (1998).\\
6. Tinney, C., Stathakis, R., Cannon, R., Galama, T.J. {\it IAU
Circ.} No. 6896 (1998).\\
7. Wieringa, M., Frail, D.A., Kulkarni, S.R., Higdon, J.L., Wark, R.
Bloom, J.S., and the BeppoSAX GRB team {\it IAU Circ.} No. 6896
(1998).\\
8. Kulkarni, S.R., Bloom, J.S., Frail, D.A., Ekers, R., Wieringa, M.,
Wark, R., Higdon, J.L. {\it IAU Circ.} No. 6903 (1998).\\
9. Soffitta, P. {\it et al.} {\it IAU Circ.} No. 6884 (1998).\\
10. Kippen, R. M. and the BATSE team. GCN Message No. 67 (1998).\\
11. Kouveliotou, C., Meegan, C.A., Fishman, G.J., Bhat, N.P., Briggs,
M.S., Koshut, T.M., Paciesas, W.S., Pendleton, G.N. Identification of
two classes of gamma-ray bursts. {\it Astrophys. J.}, {\bf 413},
L101-L104 (1993) .\\
12. Pendleton, G.N. {\it et al.} The identification of two different
spectral types of pulses in gamma-ray bursts. {\it Astrophys. J.},
{\bf 489}, 175-198 (1997).\\
13. Heise J. {\it et al.} in {\it Conf. Proc. of the 4th Huntsville
Symposium on Gamma-Ray Bursts.} (eds. Meegan, C., Preece, R., Koshut,
T. (New York: AIP) (1998).\\
14. In 't Zand, J., private communication.\\
15. Bessell, M.S. and Germany, L.M. Calibration of the MACHO
photometric system. {\it Publ. Astron. Soc. Pacif.} (in preparation)
(1998) .\\
16. Galama, T.J., Vreeswijk P.M., Pian, E., Frontera, F., Doublier,
V. and Gonzalez, J.-F. {\it IAU Circ.} No. 6895 (1998).\\
17. Duus, A. \& Newell, B. A catalog of Southern groups and clusters
of galaxies. {\it Astrophys. J. Suppl.} {\bf 35}, 209-219 (1977).\\
18. Leibundgut, B. Observations of supernovae. {\it The Lives of the
Neutron Stars}, eds. M.A. Alpar, \"U. Kiziloglu and J. van Paradijs
(NATO ASI Series; Kluwer) Series C. Vol. 450 (1995).\\
19. Schlegel, D.J., Finkbeiner, D.P., \& Davis, M. Maps of dust IR
emission for use in estimation of reddening and CMBR foregrounds. {\it
Astrophys. J.} (in the press); preprint
http://xxx.lanl.gov,astro-ph/0910327 (1998).\\
20. Patat, F. and Piemonte, A. {\it IAU Circ.} No. 6918 (1998).\\
21. Iwamoto, K. {\it et al.} A `Hypernova' Model for SN 1998bw and
Gamma-Ray Burst of 25 April 1998. {\it Nature} (submitted) (1998).\\
22. Schmidt {\it et al.} The expanding photosphere method applied to
SN1992am at cz=14600 km/s. {\it Astron. J.} {\bf 107}, 1444-1452
(1997).\\
23. Van den Bergh, S. and Tammann, G.A. Galactic and extragalactic
supernova rates. {\it Annu. Rev.  Astron. Astrophys.} {\bf 29},
363-407 (1991).\\
24. Giovanelli, R. and Haynes, M.P. Redshift survey of galaxies. {\it
Annu. Rev.  Astron. Astrophys.} {\bf 29}, 499-541 (1991).\\
25. Strom, R.G. The rate of supernovae. {\it The Lives of the Neutron
Stars}, eds. M.A. Alpar, \"U. Kiziloglu and J. van Paradijs (NATO ASI
Series; Kluwer) Series C. Vol. 450 (1995).\\
26. Pian, E., Frontera, F., Antonelli, L.A., Piro, L. GCN Message
No. 69 (1998).\\
27. Pian, E., Antonelli, L.A., Daniele, M. R.  Rebecchi, S., Torroni,
V., Gennaro, G., Feroci, M., Piro, L. GCN Message No. 61 (1998).\\
28. Galama, T.J., Vreeswijk, P.M., Groot, P.J., Stappers, B., Pian,
E., Frontera, F., Palazzi, E., Masetti, N., Nicastro, L., Feroci, M.,
Strom, R.G., Kouveliotou, C., van Paradijs, J. GCN Message No. 60
(1998).\\
29. Galama, T.J., Vreeswijk, P.M., Groot, P.J., Pian, E., Frontera,
F., Palazzi, E., Masetti, N., Nicastro, L., Feroci, M., Strom, R.G.,
Kouveliotou, C., van Paradijs, J. GCN Message No. 62 (1998).\\
30. Castro-Tirado, A. {\it et al.} {\it IAU Circ.} No. 6598 (1997).\\
31. Groot, P.J. {\it et al.} A search for optical afterglow from
GRB970828 {\it Astrophys. J.} {\bf 493}, L27-L30 (1998) .\\
32. Greiner, J., private communication.\\
33. Cagnoni, I., Della Ceca, R., Maccacaro, T. A medium survey of the
hard X-ray sky with the ASCA Gas Imaging Spectrometer: the (2--10 keV)
number counts relationship. {\it Astrophys. J.} {\bf 493}, 54-61
(1998).\\
34. Paczy\'nski, B., Are gamma-ray bursts in star-forming regions? {\it
Astrophys. J.} {\bf 494}, L45-L48 (1998) .\\
35. Landolt, A.U., UBVRI photometric standard stars in the magnitude
range 11.5 $<V<$16.0 around the celestial equator. {\it Astron. J.}
{\bf 104}, 340-376 (1992).\\

\newpage
\rm
\begin{table}
\caption[]{Times of maximum, and apparent and absolute peak magnitudes 
of SN\,1998bw. \label{tab:mag}}
\begin{tabular}{llllllllll}
\hline
 & U & B & V & R & I \\[-2mm]\hline
Date 1998 (UT) 	& May 9.6 & May 10.2  & May 12.2 & May 13.4 & May 13.8\\
Apparent mag & 13.81 $\pm$ 0.10 & 14.09 $\pm$ 0.05  & 13.62 $\pm$
0.05 & 13.61 $\pm$ 0.05 & 13.70 $\pm$ 0.05 \\
Absolute mag &-19.16 $\pm$ 0.10 &-18.88 $\pm$ 0.05 & -19.35 $\pm$ 0.05
& -19.36 $\pm$ 0.05 & -19.27 $\pm$ 0.05\\
\hline
\end{tabular}
\end{table}

\begin{figure}
\caption[]{Left panel: NTT R band image May 1.3 UT. Right panel: DSS image \label{fig:discov}}
\begin{minipage}{14.0cm}
\centerline{\psfig{figure=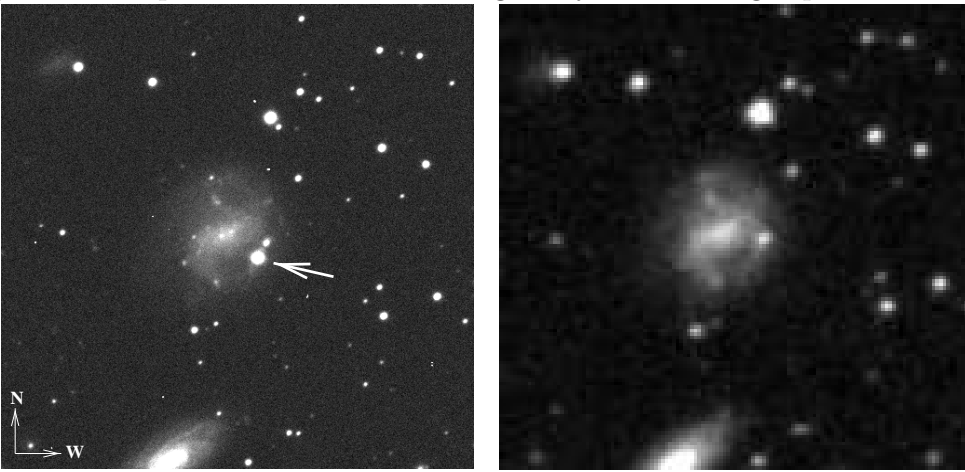,width=13cm,angle=-90}}
\end{minipage}
\end{figure} 

\begin{figure}
\caption[]{UBVRI light curves of SN\,1998bw. Time is in days since
April 25.90915 UT.  We determined a photometric (U, B, V, R and I)
calibration for a number of reference stars using NTT (May 4.4 UT) and
1.5D (May 8.3 UT) observations of the Landolt$^{35}$ fields Mark A and
SA110 (stars 496-507).  We corrected for atmospheric extinction and,
for U and B, also for a first-order colour term.  By comparison of
these two calibration nights we estimate an error of the absolute
calibration of 0.10 mag in U and 0.05 mag for B, V, R and I. The R$_M$
and B$_M$ observations have been transformed using Ref. 15.  We
consider a conservative minimum error of 0.03 mag realistic for the
differential U, B, V, R and I light curves to account for the effect
of seeing on the contribution of the underlying galaxy ($<$ 0.01 mag
for each band) and the different instruments used. \label{fig:light}}
\centerline{\psfig{figure=FIG2.ps,width=17.6cm,angle=-90}}
\end{figure}  

\end{document}